\documentclass[12pt]{article}

\usepackage[utf8]{inputenc}
\usepackage[english]{babel}
\usepackage{hyperref,times,siunitx,xcolor,mathtools,tablefootnote}
\usepackage{ae,aecompl,aeguill}
\usepackage{graphicx}
\usepackage{amsmath}
\usepackage{amsfonts}
\usepackage{amssymb}
\usepackage{bm}
\usepackage{bbm}
\usepackage{amsthm}
\usepackage[font=small,labelfont=bf]{caption}
\usepackage{subcaption}
\usepackage{cleveref}
\usepackage{color}
\usepackage{lscape}
\usepackage{sidecap}
\usepackage{pdfpages} 
\usepackage[makeroom]{cancel}
\usepackage{xr}

\DeclarePairedDelimiter{\abs}{\lvert}{\rvert}

\newenvironment{sciabstract}{
\begin{quote} \bf}
{\end{quote}}

\title{Thermo-Optically Induced Transparency\\ on a photonic chip}

\author
{Marco Clementi,$^{1\dagger}$ Simone Iadanza,$^{2,3}$ Sebastian A. Schulz,$^4$\\Giulia Urbinati,$^{1}$
Dario Gerace,$^{1}$\ Liam O'Faloain,$^{2,3}$ Matteo Galli$^{1\ddagger}$\\
\\
\normalsize{$^{1}$Dipartimento di Fisica, Università di Pavia, Via A. Bassi 6, 27100 Pavia, Italy}\\
\\
\normalsize{$^{2}$Centre for Advanced Photonics and Process Analysis, Munster Technological University,}\\ \normalsize{Rossa Ave Bishopstown, Cork T12 P928, Ireland}\\
\\
\normalsize{$^{3}$Tyndall National Institute, Lee Maltings Complex Dyke Parade,}\\ \normalsize{Cork T12 R5CP, Ireland}\\
\\
\normalsize{$^{4}$SUPA, School of Physics and Astronomy, University of St. Andrews,}\\ \normalsize{ Fife KY16 9SS, UK}\\
\\
\normalsize{$^\dagger$marco.clementi01@universitadipavia.it, $^\ddagger$matteo.galli@unipv.it}
}

\begin{document} 

\baselineskip24pt
\maketitle

\begin{sciabstract}
  Controlling the optical response of a medium through suitably tuned coherent electromagnetic fields is highly relevant in a number of potential applications, from all-optical modulators to optical storage devices. In particular, electromagnetically induced transparency (EIT) is an established phenomenon in which destructive quantum interference creates a transparency window over a narrow spectral range around an absorption line, which, in turn, allows to slow and ultimately stop light due to the anomalous refractive index dispersion. Here we report on the observation of a new form of both induced transparency and amplification of a weak probe beam in a strongly driven silicon photonic crystal resonator at room temperature. 
  The effect is based on the oscillating temperature field induced in a nonlinear optical cavity, and it reproduces many of the key features of EIT while being independent of either atomic or mechanical resonances. Such thermo-optically induced transparency will allow a versatile implementation of EIT-analogues in an integrated photonic platform, at almost arbitrary wavelength of interest, room temperature and in a practical, low cost and scalable system.
\end{sciabstract}

\section*{Introduction}

Electromagnetically induced transparency (EIT) is a coherent phenomenon in which a control laser induces transparency for a weaker probe laser in an otherwise opaque (or optically thick) medium \cite{Fleischhauer2005}. The effect was first observed in atomic vapors \cite{Boller1991}, and then exploited to slow the propagation speed of light pulses down to \SI{17}{\meter/\second} in ultracold atom clouds \cite{Hau1999}, before achieving stopped light through dynamical control of the pump laser \cite{Longdell2005}. However, EIT experiments typically require complicated set-up configurations and very low operating temperature, driving research into related and analogous effects such as coherent population oscillations (CPO) \cite{Bigelow2003}, coupled resonator structures \cite{Xu2006}, Brillouin-scattering induced transparency (BSIT) \cite{Dong2015,Kim2015} and opto-mechanically induced transparency (OMIT) \cite{Agarwal2010,Weis2010,Safavi-Naeini2011}. These effects all rely on the beating or resonant coupling between two optical modes and a set of atomic or mechanical resonances. The very narrow linewidth of the atomic or mechanical resonances leads to a rapidly varying refractive index and hence a potentially large positive or negative group index, given by $n_g=c\frac{\partial k}{\partial \omega}$. This increased group index can be used for applications such as microsecond-scale optical delay \cite{Dong2015} or improved performance of deep tissue imaging systems \cite{Zhang2012}. However, both EIT and CPO are resonant phenomena, linked to specific atomic transitions, limiting the wavelength range at which these effects can be realised. Similarly, BSIT and OMIT require structures that support acoustic or mechanical modes. While these approaches can achieve large delays (\si{\micro\second}) this comes at the cost of complex structures and device design.

Here we show a novel and general nonlinear phenomenon, based on the thermo-optic effect, displaying both induced transparency and amplification of a weak probe beam coherently coupled to a strong control field \cite{Clementi_thesis}.
The effect relies solely on the thermo-optical material nonlinearity, ubiquitous in photonic integrated devices, for which we coined the definition of ‘Thermo-Optically Induced Transparency’ (TOIT).
We hereby demonstrate that in contrast to many EIT analogues, TOIT can be realized in a CMOS-compatible platform, with a micrometer-scale footprint, and at microwatt-level power. In addition, we show that its dynamics can be controlled by just tailoring the thermal properties of the cavity used, which allows an extremely wide range of EIT bandwidths. This results in a readily accessible and easy to implement EIT analogue, in a scalable and widely available material system, which may facilitate a broader deployment of the phenomenon towards novel scientific and technological applications. 
Indeed, TOIT only requires a singly resonant cavity, with relatively moderate quality factor ($Q=38,000$ is used in this work), widely accessible in many integrated photonic plaforms and cavity geometries, yielding microsecond-scale optical delay or advance in conventional microresonators.  

We first provide an analytical and numerical model to show how thermal nonlinearities can yield induced transparency as a general effect, independent of the details of the cavity implementation. Secondly, we present an experimental demonstration in an integrated photonic crystal (PhC) cavity, coherently driven by the beating of a pump-probe beam. 
Finally, we give an outlook for potential future directions of the field, with applications such as optical and quantum memories on a photonic chip.

\section*{Results}
\subsection*{Theory}

\begin{figure}
    \centering
    \includegraphics[width=\textwidth]{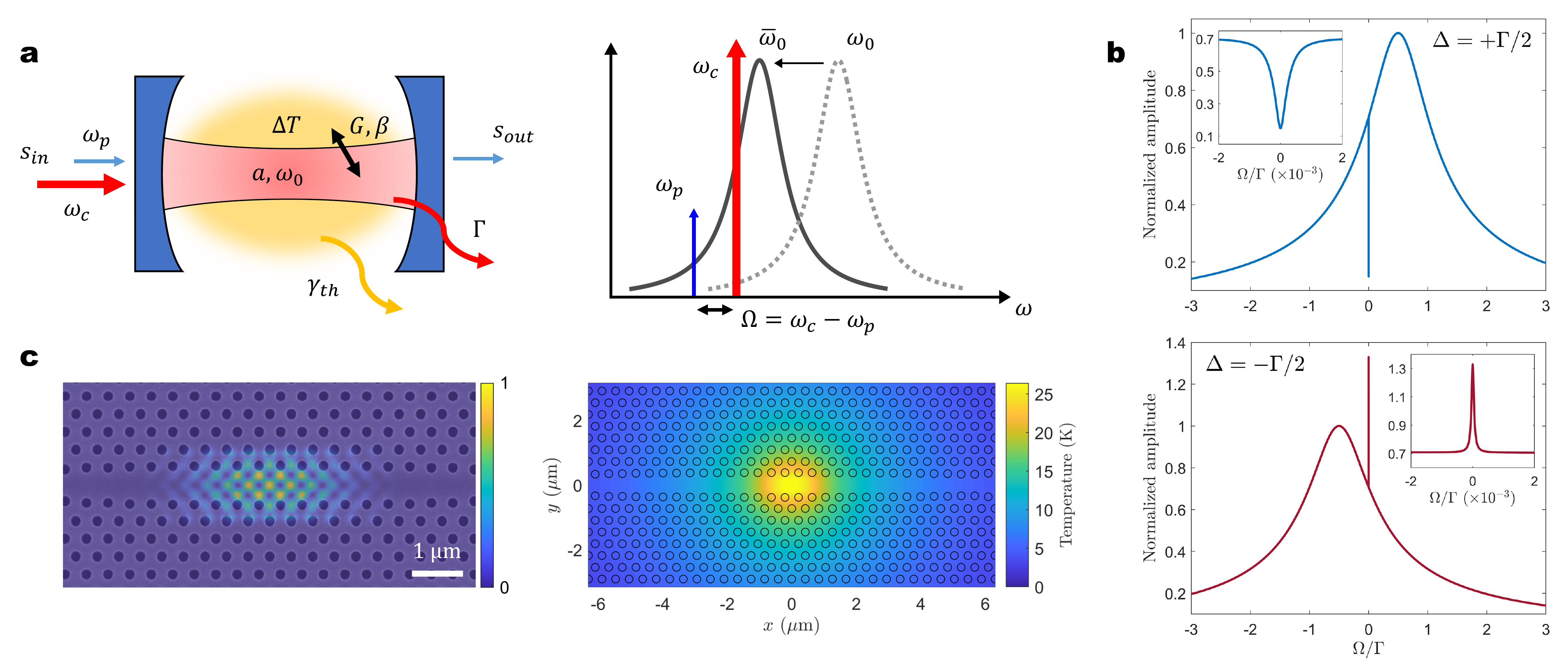}
    \caption{\textbf{control-probe of a thermo-optical cavity.} 
    \textbf{a.} (\textit{left}) Schematic of the thermo-optical cavity, showing the fields and the main physical quantities involved. 
    (\textit{right}) Thermo-optically shifted resonance (blue-detuning regime) as a static effect of the control field.
    \textbf{b.} Calculated pump-probe response from the steady state solution of the equations of motion, showing the output power oscillation amplitude $\abs{\tilde{I}}$ as a function of the pump-probe detuning frequency $\Omega$, in the amplification (red pump-cavity detuning, \textit{bottom}) and absorption (blue pump-cavity detuning, \textit{top}) regimes, respectively. The broad resonance represents the bare cavity mode, while the narrow spectral feature (inset) at zero pump-probe detuning is due to the thermo-optical nonlinearity.
    \textbf{c.} (\textit{left}) Scanning electron micrograph and simulated optical field intensity distribution (normalized) of the photonic crystal cavity used for the experiment. The optical mode localizes at the widest point of the line-defect. (\textit{right}) Simulated thermal distribution for a stationary absorbed power $P_{abs}=\SI{1}{\milli\watt}$ in correspondence of the cavity mode.  }
    \label{fig:fig1}
\end{figure}

Our experiment consists of an optical cavity displaying a self-induced frequency shift due to the thermo-optic effect, as caused by the weak optical absorption from the cavity material (Fig.~\ref{fig:fig1}a). In the case of a silicon cavity, such as the one used in this work, absorption generates free carriers, that recombine non-radiatively releasing heat. For such a cavity, the thermo-optically shifted resonance is given by $\overline{\omega}_0=\omega_0+\omega_0\alpha\Delta T$, where $\omega_0$ is the “cold” (unperturbed) cavity resonance frequency, $\alpha=\frac{1}{\omega_0}\frac{\partial\omega_0}{\partial T}$ is the refractive index temperature coefficient, and $\Delta T$ is the effective temperature offset of the cavity medium due to optical material absorption, with respect to the environment. In analogy to the formalism already introduced to interpret OMIT \cite{Weis2010}, one can define a thermo-optical coupling parameter $G=\omega_0\alpha$ such that $G\Delta T$ represents the frequency shift due to coupling of the optical mode to the thermal field.  

First, we hereby give a physical interpretation of the phenomenon. We are considering the situation in which an intense control beam induces an intracavity field amplitude $\overline{a}e^{i\omega_c t}$ oscillating at frequency $\omega_c=\omega_0+\Delta$, slightly detuned from the cold cavity resonance. The intracavity energy  $\abs{\overline{a}}^2$ gives rise to a steady  temperature increase $\overline{\Delta T}$, which in turn leads to a detuning from the “hot” (shifted) cavity resonance $\overline{\Delta}=\omega_c-\overline{\omega}_0$. In this situation, the cavity reaches the thermal equilibrium thanks to heat dissipation to the thermal bath with a rate $\gamma_{th}$\footnote{Notice that for sufficiently intense control fields, the cavity eventually enters a thermo-optical bistable regime, settling to one of the two stable equilibrium points depending on the sign and value of $\Delta$ \cite{Carmon2004}.}.

A second weak laser at a frequency $\omega_p=\omega_c-\Omega$ is used to probe the cavity response, generating a driving field that oscillates at the beating frequency $\Omega$ between the probe and the control fields. 
As a result, thermo-optical oscillations are coherently induced in the cavity, tuning the optical resonator and thereby modulating the intracavity field at exactly the same frequency $\Omega$. This leads to an interference that yields either a cancellation or an increase of the intracavity probe field depending on the relative phase between the two coherent fields, which results in a suppression or an amplification for the probe transmission. However, due to the limited thermal dissipation rate of the cavity, coherent thermo-optical oscillations may occur with appreciable amplitude only if $\Omega\sim\gamma_{th}$. This condition determines the frequency bandwidth over which the phenomenon is observed, and suggests that the narrowness of the resulting spectral feature can, in principle, be strongly decreased by reducing the thermal dissipation rate of the cavity.

More formally, we can analytically capture the essence of this behavior in terms of a simplified model in which the cavity is assumed to be uniformly heated by the thermal field, and heat is dissipated with a single decay rate from the cavity region. 

Here we consider two coupled equations for the complex intracavity field amplitude, $a(t)=\overline{a}+\delta a(t)$, and the temperature variation,  $\Delta T(t) = \overline{\Delta T}+\delta T(t)$, over the whole cavity volume (see SOM for details):

\begin{subequations}
\begin{align}
&\frac{da}{dt} =
\left(i\omega_0 - \frac{\Gamma}{2} \right)a(t) + 
i G \Delta T(t) a(t) +
\sqrt{\eta \Gamma} s_{in}(t)
\label{eq:dynamics_field_lowp}
\\
&\frac{d(\Delta T)}{dt}=
\beta \abs{a(t)}^2 - 
\gamma_{th} \Delta T (t)
\label{eq:dynamics_temperature_lowp}
\end{align}
\label{eq:dynamics_lowp}
\end{subequations}
\\
The first is the equation of motion for the driven cavity mode amplitude containing the input field, $s_{in}(t)= \overline{s}_{in} + \delta s_{in}(t)$, coupled to the cavity with efficiency $\eta$. The second equation describes the heat flow from the cavity, which is characterized by a heat capacity $C_p$, and an effective thermal conductivity $K$, such that $\gamma_{th}=K/C_p$. Here, $\beta=\Gamma_{abs}/C_p$, where $\Gamma_{abs}$  is the cavity linear absorption rate, and $\Gamma$ is the full-width at half-maximum (FWHM) of the optical cavity resonance. 
The intracavity field, the temperature variation, and the input field are given by the sum of a strong mean-field due to the control laser and a weak modulation, due to the probe laser. With this ansatz, the coupled differential equations can be solved analytically in the steady state by linearizing with respect to the modulation amplitudes, $\delta s_{in}(t)=  s_p e^{+i(\omega_c-\Omega)t}$,  $\delta a(t) = A_p^- e^{-i\Omega t} + A_p^+ e^{+i\Omega t}$ and $\delta T(t) = T e ^ {-i\Omega t} + T^* e ^ {+i\Omega t}$, respectively, $A_p^-$ and $A_p^+$ being respectively the amplitudes of the Stokes and anti-Stokes sidebands associated to the modulation of the cavity fields, and $T$ the (complex) amplitude of the temperature oscillation. The solution displays an oscillation of the output power $I(t)=\eta\Gamma\abs{a(t)}^2$ at frequency $\Omega$, which can be interpreted as the optical beating between the intracavity probe and control fields. We derived (see SOM) 
a compact and convenient expression for the oscillating component of $I(t)$:
\begin{equation}
    \tilde{I}=\left(
    1-\frac{\Gamma_{TOIT}-\gamma_{th}}
    {-i\Omega + \Gamma_{TOIT}}\right)
    \cdot
    \frac{
    2\overline{a}^* \left(\eta\Gamma\right)^{3/2} 
    }{
    i\left(\overline{\Delta}-\Omega\right) + \Gamma/2
    }\cdot
    s_p
    \label{eq:beat_analytic}
\end{equation}
\\
in which:
\begin{equation}
    \Gamma_{TOIT}=
    \gamma_{th}\left(
    1 + 
    \frac{\abs{\overline{a}}^2}{\abs{\overline{a}_b}^2}\tilde{\chi}\left(\overline{\Delta}\right)
    \right)
    \label{eq:gamma_toit}
\end{equation}
\\
and $\abs{\overline{a}_b}^2=-\frac{K\Gamma}{2G \Gamma_{abs}}$ represents the characteristic energy of the optical bistability threshold, while we defined $\tilde{\chi}\left(\overline{\Delta}\right)=\frac{4\overline{\Delta}/\Gamma}{4\overline{\Delta}^2/\Gamma^2+1}$ as a thermo-optical response function. The second multiplicative term on the right-hand side of Eq.~\eqref{eq:beat_analytic} represents the response of the bare resonator to the probe field $\delta s_{in}$, while the term enclosed in brackets describes a narrow spectral hole, or anti-hole, with half-width at half-maximum $\Gamma_{TOIT}$ and visibility $\abs{\mathcal{V}}$, such that:
\begin{equation}
    \mathcal{V}= 1- \frac{\gamma_{th}}{\Gamma_{TOIT}}
    \label{eq:visibility}
\end{equation}
 for $\Gamma_{TOIT}>0$. Both $\Gamma_{TOIT}$ and $\mathcal{V}$ depend on the pump detuning and intensity, such that $\mathcal{V}>0$ for the induced absorption regime, while $\mathcal{V}<0$ in the induced gain regime.
The resulting effect draws strict analogies with EIT and OMIT. However, the induced transparency is here due to a first-order system response, namely the thermal decay, that does not rely on any electronic or mechanical resonance, rather on the thermal field dissipated through the whole microcavity volume.

In resonators with negative thermo-optic coefficient ($\alpha<0$) a spectral hole (induced absorption) is predicted at blue detuning regime ($\overline{\Delta}>0$), with a  linearly increasing linewidth as a function of the control beam power and visibility  asymptotically approaching unity. Conversely, a spectral anti-hole (induced amplification) is expected at red-detuning regime ($\overline{\Delta}<0$), whose spectral width decreases by increasing the control power, eventually approaching zero for $\abs{\overline{a}}^2=\abs{\overline{a}_b}^2$, with diverging peak visibility.
In both conditions, the value of $\tilde{\chi}\left(\overline{\Delta}\right)$ and hence the visibility are maximized for $\overline{\Delta}=\pm\Gamma/2$. 
The calculated probe spectral response is shown in Fig.~\ref{fig:fig1}b for the blue- and red-detuning regimes.

The analytic relation between the real and imaginary parts of Eq.~\eqref{eq:beat_analytic} links the existence of a sharp spectral peak (dip) to a steep phase response for $\Omega \ll \Gamma_{TOIT}$. This is associated to large group delay (advance), whose maximum value is given by
    $\tau_g=-{\mathcal{V}}/{\gamma_{th}}$.
In analogy with the well-known phenomenology observed in bulk fast- and slow-light media \cite{Boyd2009}, the outgoing wave is advanced ($\tau_g<0$) whenever a spectral dip is observed, where the maximum group advance is here determined by the thermal response time (and hence by the thermal dissipative properties of the specific resonator structure). Conversely, arbitrarily high group delay ($\tau_g>0$) can be achieved, in principle, in the amplification regime, owing to the narrowing of the spectral feature.
It should be remarked that the bandwidth of the process is reduced under this driving condition, and thus the delay-bandwidth limit is not violated.
The maximum peak visibility and group delay achievable are in practice limited by the emergence of thermal bistability, which makes some combinations of pump power and detuning experimentally inaccessible due to the unstable nature of the thermo-optical equilibrium point \cite{Carmon2004}.

\subsection*{Experiment}

The model described in the previous section is not specific to a particular cavity design or material platform, indicating the general nature of the effect.
The  TOIT effect is most interesting in microresonators where the thermal response can vary over very different timescales, allowing a flexible trade-off between delay/advance and bandwidth.
The silicon photonics platform is very promising as there are a range of means to tailor the rate at which heat is dissipated, yielding different options for integration that provide a route to applications.
Here we experimentally demonstrate TOIT using a two-dimensional photonic crystal (PhC) cavity (Fig.~\ref{fig:fig1}c), based on a dispersion-adapted line-defect design \cite{Welna2012} and realized in a suspended silicon membrane. A simulation of the localized optical mode is shown superimposed to a micrograph of the sample, together with the associated temperature field profile generated by light absorption in the material. 

The cavity was probed by coupling the control and probe fields in a resonant scattering (RS) reflection geometry \cite{Galli2009} (see SOM). 
This allowed us to strongly suppress any spurious signals from the field reflected off the photonic crystal surface substrate and to easily isolate the field that interacted with the cavity.

The optical properties of the device ($\omega_0,\Gamma,\eta$) were investigated by linear (low-power) spectroscopy, while its thermal properties ($C_p,K$) were estimated by finite-element method (FEM). The thermo-optic coefficient $\alpha$ and the absorption rate $\Gamma_{abs}$ were estimated by temperature-dependent spectroscopy and bistability trends, respectively (see SOM). 
The extracted values are summarized in Table~\ref{table:exp_summary}.
We notice that typical values of $\gamma_{th}=K/C_p$ reported for nanoscale silicon photonic crystal cavities are as high as $\gamma_{th}/2\pi \sim \SI{1}{\mega\hertz}$ \cite{Tanabe2005,Haret2009}, due to the extremely small value of $C_p$ deriving from a diffraction-limited mode volume and the large value of $K$ provided by the surface area of the photonic crystal.

\begin{table}[tb!]
\centering
\begin{tabular}{ |c|c| } 
    \hline
    $\omega_0/2\pi$     & \SI{193.61}{\tera\hertz} \\
    \hline
    $\Gamma/2\pi$       & \SI{5.09}{\giga\hertz} \\
    \hline
    $\eta$              & \SI{0.038}{} \\ % already accounts for crossed-polarizations
    \hline
    $C_p$               & \SI{1.6e-11}{\joule/\kelvin} \\
    \hline
    $K$ & \SI{3.9e-5}{\watt/\kelvin}  \\
    \hline
    $\gamma_{th}/2\pi$ & \SI{0.38}{\mega\hertz}  \\
    \hline
    $\alpha$            & \SI{-5.85e-5}{\kelvin^{-1}} \\
    \hline
    $\Gamma_{abs}/\Gamma$ & \SI{0.64}{} \\
    \hline
\end{tabular}
\caption{Physical parameters estimated for the microcavity.}
\label{table:exp_summary}
\end{table}

In order to link the experimental results to our analytic model and quantitatively capture all the essential features of the TOIT phenomenon, the use of a single thermal decay rate assumed in the simplified model of two coupled Eqs.~\eqref{eq:dynamics_lowp} is a too crude approximation.
To fully understand the thermal response of the system, we have developed a numerical description of the heat diffusion problem within the PhC microcavity, which yields the distribution of $\Delta T$ as a function of both spatial coordinates and time \cite{Iadanza2020}. The spatially varying temperature is then linked to the optical length of the cavity, and we see that the silicon photonics platform is capable of accessing the induced transparency regime using coupled pumping powers of the order of few microwatts. However, including this full thermal description into the above rate equations would hinder the possibility to obtain an analytic solution to the problem. Therefore, in line with previous suggestions \cite{Carmon2004,Ilchenko1992}, we developed a refined version of Eqs.~\ref{eq:dynamics_lowp}, by introducing up to three interacting heat capacities, thus providing an approximation of the thermal response of the system but still allowing for an analytic solution (see SOM for details).
The use of three different heat capacities approximates correctly the description provided by our numerical simulation \cite{Iadanza2020} and can be intuitively justified by considering that heat is generated in the very small modal volume of the microcavity through optical absorption, then it rapidly diffuses through the PhC lattice around the cavity, and it is eventually dissipated to the surroundings.

\begin{figure}
    \centering
    \includegraphics[width=\textwidth]{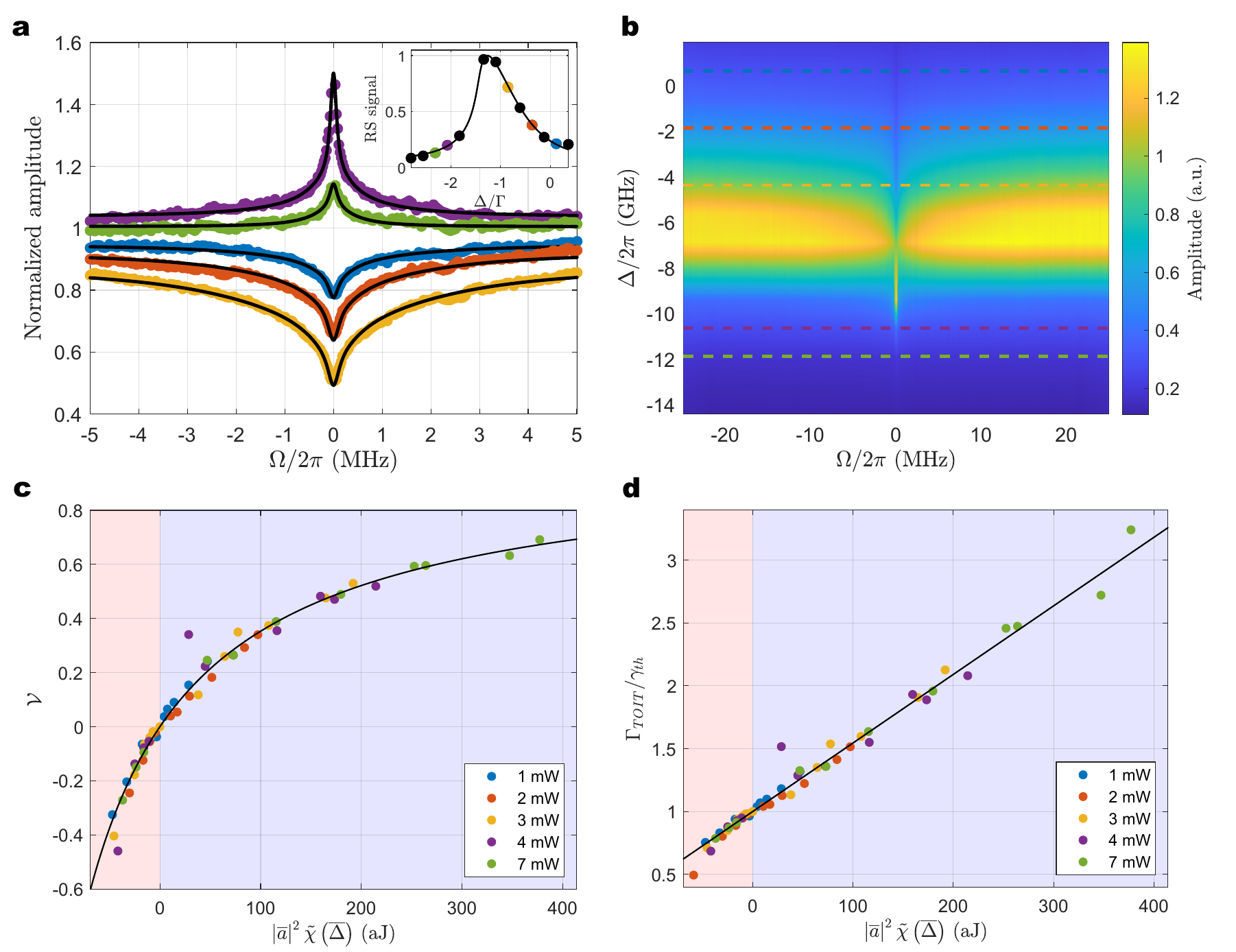}
    \caption{\textbf{Highly resolved amplitude response}. \textbf{a.} Normalized experimental spectra (dots) and model fit (black curve) for the output power oscillation amplitude $\abs{\tilde{I}}$ as a function of the control-probe detuning frequency $\Omega$ for a nominal control power $P=\SI{4}{\milli\watt}$. The traces are normalized on the average value measured at $\Omega/2\pi=\pm\SI{25}{\mega\hertz}$. (\textit{inset}) Time-averaged RS power collected during each scan (dots) and best fit with a bistable nonlinear model (black curve) as a function of the detuning $\Delta=\omega_c-\omega_0$ between the control frequency and ``cold" cavity resonance frequency. Colored points correspond to measured spectral traces and provide control-``hot" cavity detuning conditions,  $\overline{\Delta}$, and coupled power, $\abs{\overline{a}}^2$. \textbf{b.} Color map of the output power oscillation amplitude $\abs{\tilde{I}}$ as a function of the control-probe detuning frequency and control-cavity detuning . Horizontal slices correspond to experimental traces in panel a. \textbf{c.} Dip (blue region, positive values) and peak (red region, negative values) visibility as a function of the product $\abs{\overline{a}}^2\tilde{\chi}{(\overline{\Delta})}$. The visibility is calculated from a  best fit of the experimental spectra for the whole experimental dataset. The legend reports the nominal laser power for each measurement. 
    \textbf{d.} Linewidth broadening (narrowing) as a function of $\abs{\overline{a}}^2\tilde{\chi}{(\overline{\Delta})}$. The values on the vertical axis are calculated upon the data in panel c as $\Gamma_{TOIT} / \gamma_{th}=1/\left(1-\mathcal{V}\right) $.}
    
    \label{fig:fig2}
\end{figure}

To observe the TOIT phenomenon, we excited the system with a strong control laser, while the weak probe  was created by frequency shifting a fraction of the control beam through a cascade of two acousto-optic modulators. This way, we were able to scan the probe across the control within a $\Omega/2\pi$ range between -25 and 25 \si{\mega\hertz}.
The optical beat note between the control and the probe was recorded at the input and output channels of the RS apparatus by two low-noise photodetectors and used to feed the reference and signal channels of a RF lock-in amplifier, in order to reconstruct the output power (Eq.~\ref{eq:beat_analytic}) both in amplitude and phase. During each scan, the total RS signal was used to assess the detuning $\overline{\Delta}$ between the control field and the hot cavity mode (Fig.~\ref{fig:fig2}a, inset).

\begin{figure}
    \centering
    \vspace*{-3cm}
    \includegraphics[width=\textwidth]{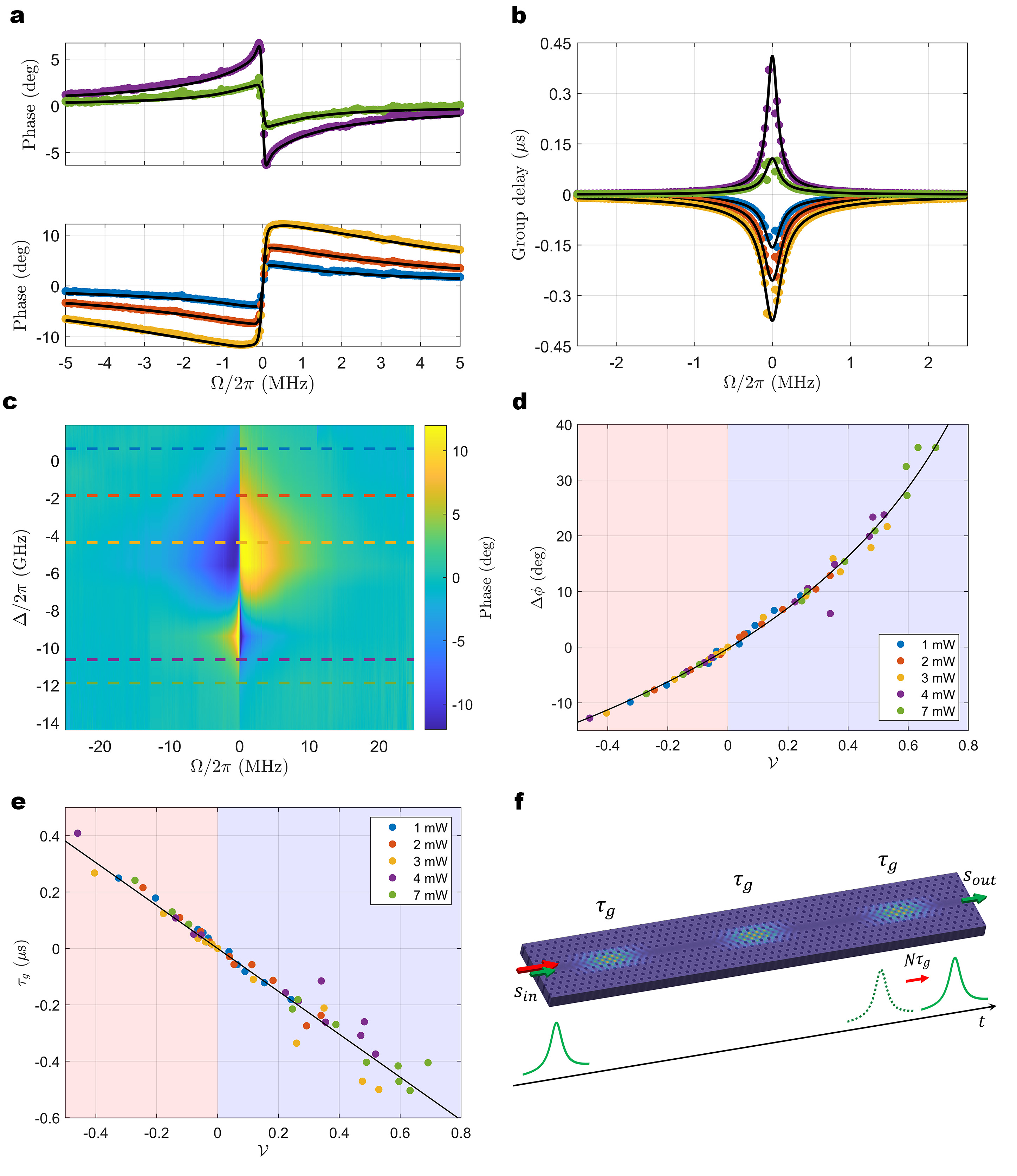}
    \caption{\textbf{Phase response.} \textbf{a.} Probe phase spectra as a function of the control-probe detuning frequency, $\Omega$. The experimental data (dots) and model fit (black curve) represent the phase trace associated to the amplitude response shown in Fig.~\ref{fig:fig2}a. Upper and lower panels correspond to red- and blue-detuning condition respectively. \textbf{b.} Group delay traces obtained by numerical differentiation of data shown in panel a. \textbf{c.} Color map of the measured phase as a function of the control-probe detuning frequency $\Omega$ and control-cavity detuning $\Delta=\omega_c-\omega_0$, associated to the measurement shown in Fig.~\ref{fig:fig2}b. Horizontal slices correspond to experimental traces in panel a. \textbf{d.} Maximum phase shift and \textbf{e.} peak group delay (advance) as a function of $\mathcal{V}$. The blue (red) region corresponds to the absorption (amplification) regime. Black curves are model fit to data. \textbf{f.} Conceptual scheme for cascaded TOIT. The group delay $\tau_g$ induced by the $N$ microcavities sums to the overall value $N\tau_g$, overcoming the time-bandwidth limit which holds for the single microresonator.
    }
    \label{fig:fig3}
\end{figure}

Each frequency scan was performed at fixed control power and detuning. For each value of control power, several spectra at different detuning conditions were acquired: an example of the overall result of such a measurement is summarized by the amplitude color scale plot in Fig.~\ref{fig:fig2}b, while the associated phase plot is shown in Fig.~\ref{fig:fig3}c. Other 4 datasets, recorded at different laser powers, are reported in SOM. 
The maximum laser power used was limited to \SI{7}{\milli\watt} in order to keep the contribution from nonlinear absorption processes (i.e., two-photon absorption and associated free-carrier absorption) negligible compared to linear absorption.

Figure~\ref{fig:fig2}a shows the recorded amplitude spectra for  \SI{4}{\milli\watt} control power (horizontal slices of Fig.~\ref{fig:fig2}b), and control frequency varying between \SI{193.593}{\tera\hertz} and \SI{193.609}{\tera\hertz}. In the presence of thermo-optic effect, the total RS power follows a characteristic sawtooth-shaped  response (Fig.~\ref{fig:fig2}a, inset).
A very narrow ($\sim$\si{\mega\hertz}) spectral feature, centered at $\Omega=0$ was clearly observed, with a crossover from absorption to amplification regime associated to the crossing of the hot resonance, which is clearly evidenced in the color scale plot.

Similarly, Fig.~\ref{fig:fig3}a shows the associated phase spectra, which exhibit a sharp dispersive response in correspondence of the center of the spectral (anti-)hole. Notice that the phase derivative is negative in the induced gain regime (upper panel) and positive in the induced absorption regime (lower panel), and reaches its maximum absolute value for $\Omega=0$. 

Model fits to the experimental data are shown as solid black curves in Fig.~\ref{fig:fig2}a and \ref{fig:fig3}a, respectively. The fitting algorithm was fed with the full lock-in trace, considering amplitude and phase response simultaneously for each experimental spectrum. The model used was based on a three heat capacities discretization, with associated decay rates  $\gamma_{1,1}/2\pi=\SI{1.77}{MHz}$, $\gamma_{2,2}/2\pi=\SI{0.58}{MHz}$ and $\gamma_{3,3}/2\pi=\SI{0.11}{MHz}$. The theoretical value predicted by the non-refined model, $\gamma_{th}/2\pi=\SI{0.38}{MHz}$ (Table~\ref{table:exp_summary}), can be interpreted as an effective parameter, which fits well within the range of the estimated decay rates. In order to further validate the adherence of the model to the cavity thermo-optical dynamics, we compared our results with temporally-resolved step-response measurements of the cold cavity as a function of the control detuning $\Delta$, as detailed in SOM.

Figure~\ref{fig:fig2}c shows the dip (peak) visibility, as defined by Eq.~\eqref{eq:visibility}, as a function of $\abs{\overline{a}}^2 \tilde{\chi}\left(\overline{\Delta}\right)$, calculated for all the acquired datasets. The experimental cavity energy $\abs{\overline{a}}^2$ and detuning $\overline{\Delta}$ were estimated from best-fit of the bistable RS traces (Fig.~\ref{fig:fig2}a, inset). From these data, we derived the ratio $\Gamma_{TOIT}/\gamma_{th}$, shown in Fig.~\ref{fig:fig2}d. Both plots show a clear trend that is correctly fitted by a hyperbola and a linear regression, respectively, yielding a characteristic bistability energy $\abs{\overline{a}_b}^2=\SI{0.2}{\femto\joule}$, in agreement with the result obtained from the measurement of the resonance shift as a function of the incident power. 

Finally, Fig.~\ref{fig:fig3}d relates the measured visibility to the maximum phase shift recorded within each scan. According to  Eq.~\eqref{eq:beat_analytic}, this is expected to follow the relation:
\begin{equation}
    \Delta\phi=2\arctan{
    \frac{1}{2}\frac{\mathcal{V}}{\sqrt{1-\mathcal{V}}}
    }
    \label{eq:phase_shift}
\end{equation}
Even in the framework of our refined model, this dependence correctly fits the data (solid line) up to a multiplicative factor ($\zeta\approx 0.58$).

To illustrate the potential of the TOIT phenomenon, we finally  discuss the slow- and fast-light effects associated. As a consequence of the driven thermo-optical oscillations, a refractive index change is produced across the extremely narrow  frequency range of the spectral feature, which implies that very large values (either positive or negative) of the group delay are expected for a probe frequency $\Omega\sim \gamma_{th}$. The measured group delay, defined as $\tau_g=-\frac{d\phi}{d\Omega}$, is shown in Fig.~\ref{fig:fig3}b as a function of the beating frequency $\Omega$. The experimental points are calculated from finite differentiation of the experimental data in Fig.~\ref{fig:fig3}a. Solid curves are best fits to our model, and clearly show a peak value for $\Omega \approx 0$. Negative values are associated to group advance, which is encountered in the absorption regime.

Fig.~\ref{fig:fig3}e shows the maximum group delay (advance) as a function of the estimated visibility $\mathcal{V}$.  
The linear relation predicted by the previously introduced expression for $\tau_g$ is clearly satisfied, as shown by a linear fit of the experimental results. This yields an effective thermal decay rate $\gamma_{th}^{eff}/2\pi=\SI{0.21}{\mega\hertz}$, in good agreement with the theoretical prediction, which corresponds to a maximum group advance of $\SI{0.76}{\micro\second}$ achievable in the blue-detuning regime.

\section*{Discussion}
First, we note that the maximum delay measured in our experiments, $\tau_g = \SI{0.41}{\micro\second}$, has a bandwidth (FWHM) of $2\pi\times 0.17$~\si{\mega\hertz}, which yields a delay-bandwidth product of $0.45$. This is associated with an amplification rate of \SI{4.4}{\dB\per\micro\second}, a result in stark contrast with other integrated photonic devices, where delay is typically associated with losses in the \si{\dB\per\nano\second} range \cite{Hu2018}. Conversely, the maximum measured advance, $-\tau_g = \SI{0.50}{\micro\second}$, is associated with a loss of \SI{5.1}{\dB} and bandwidth of $2\pi\times0.29 \si{\mega\hertz}$. This yields a delay-bandwidth product of $-0.91$ and a loss per unit time of \SI{10}{\dB\per\micro\second}.
We therefore now compare our system to other EIT analogues.
Here we establish that for BSIT and OMIT the achievable linewidth and delay or advance are given by the intrinsic linewidth/loss rate of an acoustic mode or mechanical resonance respectively. These properties may be difficult to control and improve beyond values already presented in the literature, although a remarkably high degree of control has already been achieved in actual optomechanical and phononic systems \cite{Fang2016}. Recently, an impressive demonstration of a related phenomenon was realized \cite{Ma_SciAdv_2020}, achieving millisecond delay and a similar delay-bandwidth product to our work, using a large thermally isolated table-top setup, in which heat transfer is fixed. On the other hand, the engineering of thermal energy flow from PhC cavities is a well-studied problem, where the heat flow rate can be increased through the inclusion of additional materials, for example graphene \cite{Shih2013}, or reduced by structuring the surrounding material \cite{Iadanza2020}. In Ref.~ \cite{Iadanza2020} we present a numerical model for the accurate description of the thermo-optic response of a cavity such as the one studied here under a constant pump scenario and show that structuring of the surrounding medium can engineer the heat flow out of the cavity. Applied to the TOIT effect we here show an example (see SOM) 
where support bridges are used to thermally isolate the PhC membrane from the remainder of the sample, reducing the thermal decay rate – and hence the transparency linewidth – by more than one order of magnitude. Accordingly, the theoretical delay increases by the same factor. This heat flow engineering is decoupled from the optical responses as the modifications take place sufficiently far away from the electromagnetic field of the cavity that optical properties (such as the quality factor, mode volume, resonance wavelength etc.) are not affected. This contrasts with other induced transparency phenomena, e.g. OMIT, where, for most common resonator designs, changing the mechanical resonance linewidth will typically also alter the optical properties of the cavity. While independent control of the mechanical and optical properties is possible \cite{Anetsberger2008}, OMIT requires operation in the sideband resolved regime, limiting the design freedom, e.g. when changing the optical quality factor the mechanical resonance might need to be adjusted as well to satisfy this condition.  TOIT does not require the sideband resolved regime and hence we have large freedom over the resonator design. For example, in addition to changing the thermal conduction properties of the device the heat capacity of the resonator can also be engineered, for instance by using larger (or smaller) resonant devices such as microring resonators, PhC heterostructure cavities or nanobeam cavities.

This differentiation emerges as a consequence of the different physical mechanisms underlying the two processes: while OMIT relies on the coupling of the electromagnetic field with a second-order dynamical system (i.e. a harmonic oscillator), TOIT emerges from the interaction between light and a system obeying a first-order dynamical response. Therefore, the TOIT spectral response is determined by the thermal dissipation rate, while in the case of OMIT, the presence of a characteristic resonant frequency also plays a fundamental role, determining the detuning between the pump and spectral hole/anti-hole, and hence the requirement for sideband resolved regime.

Lastly, we address an issue inherent to all induced transparency phenomena. Induced transparency is a resonant process and as such, the theoretical delay-bandwidth product is limited to $\tau_g\Gamma_{EIT}=1$. This can be overcome in conventional resonant systems by either time-dependent control of the cavity \cite{Xu2007}, or by cascading multiple cavities, and the delay can become arbitrarily high, in principle \cite{Schulz2010,Cardea2020}. Similar approaches  could be applied to induced transparency systems. However, the technological challenges associated with cascading multiple systems are prohibitive for most implementations. In the case of OMIT, as an example, this would require the exact matching of both the mechanical and optical resonance frequencies of multiple high Q-factor cavities. On the other hand, for TOIT only the optical resonance wavelengths need to be matched (within the cavity linewidth), and the possibility to use moderate Q-factor cavities - 38,000 in this work vs 500 million in a table-top Fabry-P\'erot cavity \cite{Ma_SciAdv_2020} – greatly simplifies this task, which can be easily and accurately performed after the device fabrication \cite{Clementi2019a}. Therefore, as sketched in Fig.~\ref{fig:fig3}f, one could easily devise multiple integrated cavities cascaded along a waveguide, which would allow to achieve larger delays for broader signals than what is achievable with a single resonant or induced transparency system.

%\section{Conclusions}

We have reported on the experimental evidence and a detailed theoretical understanding of a new form of electromagnetically induced transparency occurring in a silicon integrated photonic platform at room temperature. The new phenomenology relies on the slow thermo-optical response of a photonic crystal resonator, and manifests itself as a narrow spectral window of anomalous dispersion induced by control-probe interference within the cavity, which yields group delays or advances in the order of microseconds. The delay and bandwidth of this system can be tailored to specific application requirements by controlling the thermal properties of the cavity, independent of its optical properties, or by cascading multiple cavities in a row, each exhibiting a TOIT window. 
The TOIT phenomenon is fully compatible with chip integration using CMOS fabrication and operates at room temperature. The modest cavity Q-factors and low pump powers used make the TOIT approach to induced transparency and amplification easy to implement and scalable, providing a route to the widespread use of large and controllable optical delays in integrated photonics, with applications ranging from delay lines to optical quantum memories.

\appendix
\section*{Materials and methods}
\subsection*{Cavity design and fabrication}
%% see supplementary

The PhC cavity employed in this work is based on a dispersion adapted (DA) design \cite{Welna2012}, in which the width of a line defect in a triangular lattice of air holes is modulated to obtain wavelength scale confinement of the electromagnetic field along the propagation direction. The DA PhC cavities are simulated by finite-difference time-domain (3D-FDTD commercial software from Lumerical Solutions). The original cavity design is modified by introducing a far-field optimization strategy \cite{Portalupi2010a} to improve coupling efficiency with free-space excitation. The thermal properties of the DA PhC cavity were simulated by a finite element method (FEM) using the same commercial software. The results are shown in Fig.~1b. The thermal conductance and heat capacity of the PhC cavity are then extracted from these simulations (see Table~\ref{table:exp_summary}).

The device measured in this work was fabricated by exposing a ZEP 520A resist by means of electron-beam lithography at \SI{30}{keV}, and transferring the pattern of circular holes with approximately \SI{240}{\nano\meter} diameter and \SI{420}{\nano\meter} lattice constant onto the \SI{220}{\nano\meter} thick silicon layer of a silicon-on-insulator (SOI) wafer with a \SI{2}{\micro\meter} thick buried oxide layer through reactive ion etching using CHF$_3$ and SF$_6$. The buried oxide layer underneath the PhC region was removed using a liquid hydrofluoric (HF) acid etch.

\subsection*{Experimental techniques}
The pump and probe signal is generated as follows.
A continuous-wave tunable light source centered at \SI{1550}{nm} is sent to a fiber-coupled beam-splitter (FBS). At the first output arm, light propagates unperturbed, while at the second a cascade of two acousto-optic modulators (AOM) first shifts the light frequency by $\Omega$. The two arms are then recombined by a second FBS. The output signal thus generated is then composed of a pump field at frequency $\omega_c$ and a weak probe at frequency $\omega_p$, with intensity ratio 10:1, and it is used to feed the RS apparatus.
This is composed by a free-space cross-polarization spectroscopy arrangement \cite{Galli2009}, where the collimated light passes first through a linear polarizer and then through a beam-splitter cube. The transmitted Gaussian beam is then focused on the sample by a microscope objective to a diffraction limited spot, the beam numerical aperture being optimized to match the one of the cavity mode. The sample orientation is chosen to obtain a \SI{45}{\degree} orientation of the far-field polarization with respect to the one of the input beam. The output signal is collected along the same excitation path: the BS redirects part of it to a second linear polarizer in order to suppress any spurious signal from the sample substrate, effectively isolating only the fraction of the field which is resonantly coupled to the resonator and backscattered in the crossed polarization.
The input and output signals of the RS setup are photodetected and amplified, and the resulting electrical signals are fed to a RF lock-in amplifier, respectively as reference and signal input. The phase of the optical beat on the input signal is thus taken as a reference for the output, thus providing an amplitude and phase readout for the quantity \eqref{eq:beat_analytic}.
\subsection*{Data analysis and modelling}
%% numerical model
{\bf Numerical simulation of thermo-optic cavity response.}\\
The thermal behavior of the optically pumped silicon microcavity is governed by the heat generation associated to optical absorption via, e.g., either defect states or absorptive nonlinear processes such as two photon absorption and free carrier absorption, as well as by heat dissipation through the material, which depends on heat diffusion through the PhC lattice, the unpatterned silicon, and the substrate. To account for all these heat generation and diffusion mechanisms, we have developed and numerically solved a model of nonlinearly coupled rate equations, which provides a rigorous description of the time-dependent behavior as a function of the pump power coupled to the single mode resonator. A full description of this model, its numerical solution, as well as an experimental validation of the results in the different pumping regimes has been recently published in Ref.~ \cite{Iadanza2020}, to which we refer for details. \\
{\bf Analytic model with multiple thermal decay rates}\\
While the heat diffusion from the microcavity can only be accurately described through an infinite number of thermal decay times, a refined version of the simplified model presented in the main text can be formulated by considering $n$ concentric regions with varying and discretized heat capacities (and hence thermal decay times) from the inner to the outer shell. This is essentially accomplished by adding $n$ coupled equations for $\Delta T_i$ thermal gradients in each of the concentric shell ($i=1,\ldots,n$), and thus generalizing the model in Eqs.~\ref{eq:dynamics_lowp} to $n+1$ coupled equations, which is described in detail in SOM. 
While this generalization could bring arbitrary highly accurate solutions, we explicitly keep up to 3 thermal diffusion times (i.e., defining the rates $\gamma_{1,1}$, $\gamma_{2,2}$, and $\gamma_{3,3}$ in the main text), which still allows for a closed analytic expression of the steady state response of the system (see SOM). 
This is shown to be sufficient to quantitatively fit the experimentally measured lineshapes of the induced probe transparency and amplification in Figs.~2 and 3. 
The use of a minimum of three decay rates is justified a posteriori by fitting the numerically solved cavity response with an infinite number of decay rates (see previous paragraph).

\section*{Acknowledgments}
M.C., G.U., D.G., and M.G. acknowledge the Horizon 2020 Framework Programme (H2020) through the QuantERA ERA-NET Cofund in Quantum Technologies, project CUSPIDOR, cofunded by Ministero dell’Istruzione, dell’Università e della Ricerca (MIUR), and MIUR through the “Dipartimenti di Eccellenza Program (2018-2022)”, Dipartimento di Fisica, Università di Pavia.
S.I. and L.O. acknowledge funding from the Science Foundation Ireland (17/QERA/3472, 12/RC/2276 P2) and in part by the European Research Council Starting Grant 337508 (DANCER) and under Grant 780240 (REDFINCH).

\section*{Competing interests}
The authors declare no competing interests.

\clearpage

\end{document}